\documentclass{elsart}
\usepackage{mathrsfs}
\usepackage{amssymb}
\usepackage{amsmath}
\usepackage{amsfonts}
\usepackage{graphicx}           
\usepackage{rotating}    
\usepackage{dcolumn}
\usepackage{epsfig} 
\usepackage[all]{xy}

\newtheorem{propo}{Proposition}
\newtheorem{defi}{Definition}

\begin{document}

\begin{frontmatter}

\title{Non-transitive maps in phase synchronization}
\author{M.  S. Baptista$^{1,2}$,} \ead{murilo@agnld.uni-potsdam.de, 
Phone: +49\ 30\ 1977-1432, FAX: +49\ 30\ 1977-1142}  
\author{T.  Pereira$^{1,2}$,}
\author{J. C.  Sartorelli$^2$,} 
\author{I. L. Caldas$^2$,}
\author{and J. Kurths$^1$,}
\address{$^{1}$ Universit{\"a}t Potsdam, Institut f{\"u}r Physik
Am Neuen Palais 10, D-14469 Potsdam, Deutschland} 
\address{$^2$ 
Instituto    de  F{\'\i}sica,  Universidade  de  S{\~a}o
Paulo\\ Caixa Postal 66318, 05315-970 S{\~a}o Paulo, SP, Brasil} 

\date{\today} 

\begin{abstract}
  Concepts from Ergodic Theory are used to describe the existence of
  special non-transitive maps in attractors of phase synchronous
  chaotic oscillators. In particular, it is shown that for a class of
  phase-coherent oscillators, these special maps imply phase
  synchronization. We illustrate these ideas in the sinusoidally
  forced Chua's circuit and two coupled R{\"o}ssler oscillators.
  Furthermore, these results are extended to other coupled chaotic systems. In
  addition, a phase for a chaotic attractor is defined from the
  tangent vector of the flow. Finally, it is discussed how these maps
  can be used to a real-time
  detection of phase synchronization in experimental systems.\\
\end{abstract}

\end{frontmatter}

\maketitle 

\section{Introduction}

Coupled chaotic systems are recently calling much attention due to the
verification that they may be useful to the understanding of natural
systems in a variety of fields as in ecology \cite{blasius}, in
neuroscience \cite{reynaldo,juergen}, in economy \cite{economy}, and
in lasers \cite{imaculada,laser}.  It has been verified that despite
of the higher dimensionality of a coupled chaotic system, the coupling
among the elements might make them to synchronize
\cite{reviews,strogatz}, reducing the dynamics of the system to a few
degrees of freedom.

In this work, we focus our attention in the phenomenon of Phase
Synchronization (PS), which describes the appearance of a phase
synchronous behavior between two interacting chaotic systems
\cite{rosenblum}, i.e., given two chaotic systems, their phase
difference remains bounded, despite of the fact that their amplitudes
may be uncorrelated. This phenomenon is particularly interesting since
it can arise from a very small coupling strength. Its presence was
reported in a variety of experimental systems. It was 
experimentally demonstrated in electronic circuits \cite{parlitz}, and
latter in electrochemical oscillators \cite{hudson}. It was found
in plasma \cite{epa}, in the Chua's circuit \cite{mu_chua}, and there
were also found evidences of phase synchronization in communication
processes in the Human brain \cite{fell:2002,mormann:2003} .

To detect PS in a real-time experiment, one has to measure the phase
of the chaotic trajectory \cite{phase_definitions}. However, the phase
is not always an easily accessible information. To overcome this
difficulty, it is important to understand  fundamental
properties of phase synchronous systems, that could be experimentally
easily verified. For chaotic systems that are phase synchronized with
a periodic forcing \cite{perturbed}, it was reported that a
stroboscopic map of the trajectory was a subset of this attractor and
occupies only partially the region delimitated by a projection of
the attractor.  This property was used to detect in a real-time 
experiment phase synchronization between the Chua's circuit and a sinusoidal forcing
\cite{mu_chua}.

This approach of detecting phase synchronization through the
stroboscopic map can be extended for coupled chaotic oscillators, in a
formal way.  The stroboscopic map is generalized to the {\bf
  Conditional Poincar\'e Map}. Given two oscillators,
at least one being chaotic, the conditional Poincar\'e map is
constructed by collecting points in one oscillator at the moment at which 
some event happens in the other one.  If the set of discrete points
generated by this conditional map does not visit any arbitrary region of a
 especial projection of the chaotic attractor, we call this set a {\bf
  P-set}.  This property of the conditional Poincar\'e map is called
non-transitivity \cite{transitivity}, i.e., an initial condition under
the conditional Poincar\'e map does not visit everywhere in a
subspace of the attractor. Alike the stroboscopic maps of oscillators
that are in phase synchrony with a forcing, the conditional Poincar\'e
maps of coupled chaotic oscillators, in PS, also only partially occupy a
projection of the attractor.

In this work, we show how the conditional Poincar\'e map can be used 
to detect PS, without actually having to measure the phase. For
phase-coherent oscillators, a special type of P-set, that
we call PS-set (Phase Synchronization set), exists. Conversely,
its existence also implies PS. We illustrate our findings and ideas with
numerical and experimental analyzes in the forced Chua's circuit, and
the coupled R\"osller oscillator \cite{rossler}.

Further, we extend these results to non-phase coherent attractors.
Finally, we also introduce a phase of a chaotic trajectory to be a
quantity related to the amount of rotation of the tangent vector.
This definition can be used to chaotic attractors, independently whether
they have phase-coherent or non phase-coherent dynamics.

This work is organized as follows: in Sec.  \ref{sec:phase}, we define
a way to measure the phase of a chaotic flow, and discuss the relation
between the average returning time and the average angular frequency.
 In Sec.  \ref{sec:PS}, we discuss
the conditions for PS states and, in Sec.
\ref{sec:chua}, we describe the phenomenon of PS in the forced Chua's
Circuit. We introduce the notion of a conditional Poincar\'e map in
Sec.  \ref{CPM} and the P-sets (as well as the PS-sets) in Sec.
\ref{P-sets}.  In Sec.  \ref{sec:bsc}, we show how PS can be found by
the detection of these sets in the forced Chua's Circuit and in Sec.
\ref{sec:bsr}, for the coupled R\"ossler oscillator.  Further, in Sec.
\ref{extension}, we discuss the extension of these ideas to a class of
non-coherent oscillators.  In Sec.  \ref{sec:cr}, we make some remarks
and the conclusions of this work.  In Appendix A, we formally
introduce the conditional Poincar\'e map and the P-set, and in
Appendix B, we show that for coherent dynamics the PS-sets
exist if, and only if, there is PS. In other words, PS implies PS-sets
and vice-versa.

\section{Phase, frequency and average returning time of a chaotic 
  attractor} \label{sec:phase}

The phase of  a chaotic attractor in a projection $j$ (a
subspace) is defined to be the amount of rotation of the tangent 
vector in this projection, and can be 
represented by an integral function of the type

\begin{equation}
\phi_j(t) = \int_0^t \left|\frac{d \theta(t^{'})}{d t^{'}} \right| dt^{'}
\label{fase}
\end{equation}
\noindent
with $d \theta(t)$ being an infinitesimal displacement of the tangent
vector of the flow, from time $t$ to time $t+dt$, and $dt \rightarrow
0$. Note that in Eq. (\ref{fase}), we are measuring the amount of
rotation, per unit time, of a projection of the tangent vector of the
flow, on the same subspace $j$ where the phase is defined. We call
this subspace $\mathcal{P}_j$. The attractor $\mathcal{X}$, projected on the subspace 
$\mathcal{P}_j$ is regarded as $\mathcal{X}_j$. 
The instantaneous angular frequency of
the trajectory in $\mathcal{X}_j$, named $W_j$ is given
by $ \frac{d \phi_j}{dt}$. So, from Eq.  (\ref{fase}), $W_j =
\left|\frac{d \theta}{dt} \right|$ and, the average angular frequency
$\langle W_j\rangle$ is
\begin{equation}
\langle W_j \rangle =\lim_{t \rightarrow \infty} \langle \frac{d\phi_j}{dt} \rangle, 
\label{natural_frequence_2}
\end{equation}
\noindent
$ \langle - \rangle$ represents the average. Equation
(\ref{natural_frequence_2}) can be put into the form $\langle W_j
\rangle$ = $\frac{\phi_j(t)}{t}$.

Whenever a Poincar\'e section can be defined, the average period of
the chaotic attractor on the subspace $\mathcal{P}_j$ is calculated by
\begin{equation}
\langle T_j \rangle=
\frac{\sum_{i=0}^{N} \Delta \tau^i_j }{N},
\label{average_period}
\end{equation}
\noindent
where $\Delta \tau^i_j = \tau^{i}_j - \tau^{i-1}_j$, and $\tau^{i}_j$
represents the time at which the trajectory in the subspace
$\mathcal{P}_j$ makes the $i$-th crossing with this Poincar\'e
section.

We introduce $\langle \Delta \phi_j \rangle$ to be the average
displacement of the phase for a typical period as
\begin{equation}
\langle \Delta \phi_j \rangle = \frac{\phi_j(N)}{N}, 
\end{equation}
\noindent 
with $\phi_j(N)$ being the phase associated to the subspace
$\mathcal{P}_j$ at the moment that the $N$-th crossing between the
trajectory and the Poincar\'e section happens.

Thus, we can put Eq. (\ref{natural_frequence_2}) as 
\begin{equation}
\langle W_j \rangle = 
\frac{\langle \Delta \phi_j \rangle}{\langle T_j \rangle}. 
\label{natural_frequence_1}
\end{equation}
\noindent
For the forced Chua's circuit, the subspace $\mathcal{P}_1$ is defined
by a suitable projection of the circuit variable.  We have that
$\langle \Delta \phi_1 \rangle = 2\pi$.  So, Eq.
(\ref{natural_frequence_1}) can be written as $\langle W_1 \rangle =
\frac{2\pi}{\langle T_1 \rangle}$.  For the coupled R\"ossler
oscillator, the quantities in Eq. (\ref{natural_frequence_1}) can be
calculated in two subspaces, one associated to the variables of one
R\"ossler, the subspace $\mathcal{P}_1$, and the subspace
$\mathcal{P}_2$, associated to the variables of the other R\"ossler
system. As shown in \cite{murilo_potsdam}, $\langle \Delta \phi_1
\rangle$ might slightly differs from $2\pi$, and thus, $\langle W_j
\rangle = \frac{\langle \Delta \phi_j \rangle }{\langle T_j \rangle}$.

So, Eqs.  (\ref{natural_frequence_2}) and (\ref{natural_frequence_1})
relate the average period, the average angular frequency, and the
phase of a chaotic trajectory. This shows that the average period
(recurrence) and the average angular frequency are intimately
connected in phase coherent chaotic systems, and both these quantities
can be calculated from the phase.

\section{Phase Synchronization}\label{sec:PS}

Having defined phase, PS exists whenever the
following condition is satisfied {\small

\begin{equation}
|\phi _{1}(t)-r\phi _{2}(t)|< \langle
  \Delta \phi_1 \rangle. \label{PS} 
\end{equation}}
The minimal bound for the phase difference, $\langle \Delta \phi_1
\rangle$, in terms of the phase as defined by Eq. (\ref{PS}), 
was theoretically estimated in Ref.  \cite{murilo_potsdam}.
Equation (\ref{PS}) means that the phase difference between the two
coupled systems is always bounded, and $r$  
is a rational constant \cite{cita_r}.  

Also,
\begin{equation}
\langle         W_1     \rangle-   r      \langle
W_2 \rangle = 0.
\label{NC}
\end{equation}
\noindent
In this work, we will consider cases for $r$=1. Otherwise, a simple
change of variables could eliminate this constant from Eq. (\ref{NC}).
For the forced Chua's circuit $\langle W_2 \rangle = \omega$, with
$\omega=2\pi f$, representing the angular frequency of the forcing.
There is PS, if $\langle W_1 \rangle$=$\omega$. Therefore, $\langle
T_1 \rangle$=$1/f$.  For the coupled R\"ossler, if PS exists, $\langle
W_1 \rangle$ = $\langle W_2 \rangle$, $\langle R_1 \rangle$=$\langle
R_2 \rangle$, and $\langle \Delta \phi_1 \rangle$=$\langle \Delta
\phi_2 \rangle$, and therefore, we could have in the right term of Eq.
(\ref{PS}) $\langle \Delta \phi_2 \rangle$, instead of $\langle \Delta
\phi_1 \rangle$.

\section{The sinusoidally forced Chua's  circuit}\label{sec:chua}

The circuit is represented by: 

\begin{eqnarray}
C_{1} \frac{dX_{1}}{dt}=g(X_{2}-X_{1})-i_{NL}\\
C_{2}\frac{dX_{2}}{dt}=g(X_{1}-X_{2})+X_{3}\\
L\frac{dX_{3}}{dt}=-X_{2}-V\sin {(\omega t)}
\end{eqnarray}
\noindent
where $X_{1}$, $X_{2}$, and $X_{3}$ represent, respectively, the
tension across two capacitors and the current through the inductor
(See \cite{mu_chua} for more details), $\omega$ and $V$ are the
angular frequency and the amplitude of the forcing, respectively.
The piecewise-linear function, $i_{NL}$, is given by:

\begin{equation}
i_{NL}=m_0 X_1 + 0.5(m_1-m_0)[|X_1+B_p| - |X_1-B_p|]
\end{equation}
\noindent
where we have chosen the parameters $C_1$=0.1, $g$=0.574, $C_2$=1,
$L$=1/6, $m_0$=-0.5, and $m_1$=-0.8, such that we obtain a
R\"ossler-type attractor, for $V$=0.

\begin{figure}[!h]
\centerline{\hbox{\psfig{file=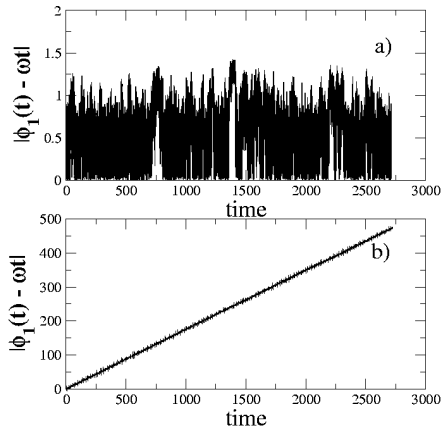,height=7cm}}} 
\caption{(a) Phase difference is always smaller than 2$\pi$, so
  PS exists between the circuit and the forcing, for $f$=0.2801 and
  $V$=0.0015. (b) PS is not present and the phase difference grows bigger
  than 2$\pi$.}
\label{PS_fig6}
\end{figure}

To calculate the phase of the chaotic trajectory, we first define the
subspace $\mathcal{P}_1$ to be given by the pair of variables $(X_1,
X_2)$, and then, we use Eq. (\ref{fase}).  In Fig.  \ref{PS_fig6}, we
show the difference between the phase of the chaotic circuit [as
calculated by Eq.  (\ref{fase})] and the phase of the forcing, $\omega
t$. In (a), the phase difference is bounded and the average period of
the chaotic attractor, defined as the average recurrence time of
trajectories that cross the section $X_2=0$, is equal to $\langle T_1
\rangle$=3.57015, which is equal to 1/f, since $f$=0.2801.  The
average angular frequency can be calculated using Eq.
(\ref{natural_frequence_1}), which gives us $\langle W_1
\rangle$=1.75992. Or, from Eq.  (\ref{natural_frequence_2}), which
gives us $\langle W_1 \rangle$=1.75992. Note that the average growing
of the phase [calculated by Eq. (\ref{fase})] for a typical average
period is $6.28318\ldots$ which is $2\pi$.  In Fig.  \ref{PS_fig6}(b),
we have that $\langle T_1 \rangle$=3.57006 which is different from
$1/f$, since $f$=0.279. So, in (b) there is no PS, and consequently
Inequality (\ref{PS}) is not satisfied.

In Ref.  \cite{mu_chua}, we detected $PS$ experimentally in the forced
Chua's circuit, whenever stroboscopic maps could be constructed for a
time interval equal to $\Delta \tau_1=\frac{1}{f}$, such that this
map, projected into the same subspace considered to calculate the
phase, does not occupy the region occupied by the attractor projected on
the same subspace.

To understand this technique, we assume that $\phi_1(t)$, the phase of
the chaotic trajectory, is the angle (on the lift) described by the
vector position of this trajectory (R\"ossler-Like attractor), and
$\phi_2(t)$=$\omega t$ the phase of the forcing.  If Eq.  (\ref{PS})
is satisfied at any time, then it is satisfied at multiples of the
period of the forcing $\tau_1^i=\frac{i}{f}$.  So, we get
$|\phi_1(\tau_1^i)-2\pi i|<2\pi$, which means that a stroboscopic map
has to be concentrated in an angular section smaller than $2\pi$. The
stroboscopic map, which is already a subset of the chaotic flow,
projected into the same subspace considered to calculate the phase,
does not occupy the whole region occupied by the attractor projected
in this same subspace. Another property of the stroboscopic map is
that points in it are mapped into it by looking at the trajectory
after a time interval given by $\Delta \tau_1$, so, it is a subset
that is recurrent to itself.

\begin{figure}[!h]
\centerline{\hbox{\psfig{file=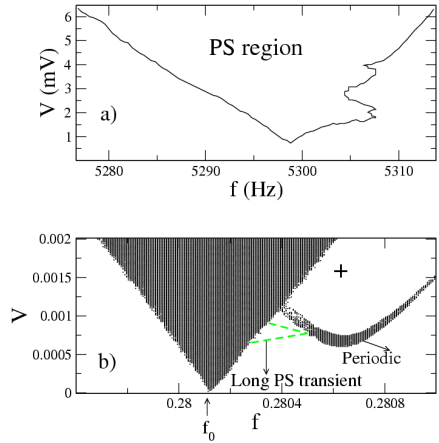,height=7cm}}} 
\caption{(a)  Experimental  $PS$ parameter   space. (b)  Simulated $PS$
  parameter space. Black points represent parameters for which Eq.
  (\ref{PS}) is satisfied for 120,000 crossings of the trajectory at
  $X_2$=0.  In both figures, the horizontal axis represents the
  forcing frequency $f$ and the vertical axis its amplitude $V$.
  Variables in (b) are dimensionless and $f_0$ is the main frequency
  of the non-forced circuit.}
\label{PS_fig1}
\end{figure}

Using this technique, and for the same parameters as \cite{mu_chua},
we show in Fig.  \ref{PS_fig1}(a) the experimental synchronization
region for the forced Chua's circuit in the parameter space $V \times
f$. The triangular shaped region represents parameters for which the
stroboscopic map has the points concentrated in an angular section
smaller than 2$\pi$. The bump at the bottom right side of the PS
region is due to non-synchronous states that present a long bounded
phase difference. By a typical time interval within which the
experiment is realized, which was of the order of 40,000 cycles, the
system seemed to be phase synchronized, i.e., localized stroboscopic
maps were found.  In fact, we have detected these maps even for
observation times corresponding to 150,000 cycles, in the region of
the bump. 

In a short, the bump region is an extended structure in the
circuit parameter space, that presents intermittent behavior in the
phase difference \cite{intermittency}, but with a long laminar regime,
even for parameters far away from the border between the PS and the
non PS region.  This intermittency differs from the usual one,
observed in the transition to PS, in the fact that this latter happens
very close to the border between the PS and the non PS region.  The
reason for this intermittency is due to the presence of a periodic
window, close to the region of the bump.

Simulation is shown in Fig.  \ref{PS_fig1}(b), where black points
represent perturbing parameters for which Eq.  (\ref{PS}) is
satisfied, with $\phi_1$ defined in Eq.  (\ref{fase}).  One sees that
the $PS$ region resembles a triangle.  The triangular shaped region,
denoted by the light gray dashed line, represents the region where the
system is not phase synchronized, but the phase difference remains
bounded for a long time interval, that might be longer than 100,000
cycles of the systems.  So, we reproduce numerically the same atypical
intermittency observed experimentally, i.e., long laminar regime in
the phase difference, for parameter regions away from the border
between PS and non PS states. This happens associated with a periodic
window, as the one shown in Fig. \ref{PS_fig1}(b).

\begin{figure}[!h]
\centerline{\hbox{\psfig{file=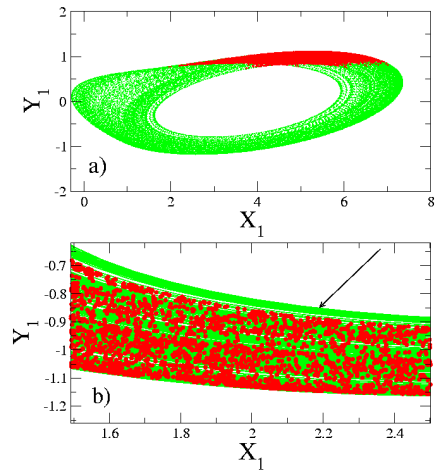,height=7cm}}} 
\caption{A projection of the attractor in light gray, and its 
  stroboscopic map in dark gray.  The parameters are $f$=0.2801 and
  $V$=0.0015, in (a), and $f$=0.28063 and $V$=0.0016, in (b).}
\label{PS_fig7}
\end{figure}

The shape of the synchronization region in Fig.  \ref{PS_fig1}(b) is
equivalent to the region in the experiment, constructed by detecting
the stroboscopic maps contained within a small angular section.  This
proposes an equivalence between the existence of a recurrent subset and
the verification of Eq. (\ref{PS}).  Inside the synchronization
region, a stroboscopic map appears like in Fig. \ref{PS_fig7}(a),
where the light gray points represent the attractor, and the dark
filled circles, the map. Outside of the $PS$ region, there are
parameter sets for which the stroboscopic maps do not occupy the
region occupied by the attractor, at the projection in which the phase
is calculated. As one sees in Fig.  \ref{PS_fig7}(b) [for the
parameters represented by the plus symbol in Fig.  \ref{PS_fig1}(b)] there
is a region of the projected attractor (pointed by the arrow) for
which the stroboscopic map never visits.

The difference between the stroboscopic map that appears while there
is PS in (a), and the stroboscopic map that appears while there is not
PS in (b), is that points of the stroboscopic map in (a) are all
concentrated in an angular section of the attractor.  As it will be
further classified, the stroboscopic map in (a) is a PS-set and in (b)
is a P-set.

\begin{figure}[!h]
\centerline{\hbox{\psfig{file=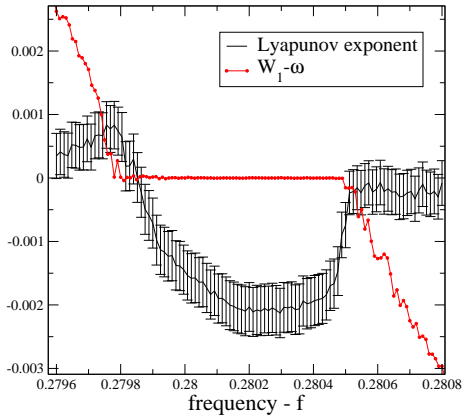,height=7cm}}} 
\caption{The Lyapunov exponent (and the error bars) 
  associated to PS in black, and the quantity $\langle W_1 \rangle
  -\omega$ in gray, for a fixed amplitude of $V$=0.0015 and a varying
  frequency.}
\label{PS_fig8}
\end{figure}

As a way to better characterize the PS phenomena in the Chua's
circuit, we calculate the Lyapunov exponents. As usually expected, the
transition to PS is associated to one of the exponents becoming
smaller than zero.  Since there is already one exponent that is
smaller than zero, PS induces the creation of a second stable
direction in the Chua's circuit. Previous to the transition to phase
synchronization, this exponent was zero. In Fig.  \ref{PS_fig8}, we
show the exponent (and the error bars) associated to PS in black and
the quantity $\langle W_1 \rangle -\omega$ (in gray, for a fixed
amplitude of $V$=0.0015) with respect to the frequency. In the region that the
exponent becomes smaller than zero, $\langle W_1 \rangle$=$\omega$,
satisfying Eq. (\ref{NC}).

\section{Conditional Poincar\'e Map}\label{CPM}

The finding of maps of the attractor that appear as localized
structures implies PS. The conditional Poincar\'e map introduced in
this chapter as a generalization of the stroboscopic map, is an
efficient way of revealing the existence of such special mappings.

The stroboscopic map technique defined for periodically driven chaotic
systems was explained in the previous sections.  To extent the idea of
stroboscopic map in coupled chaotic oscillators we came up with the
conditional Poincar\'e map, which is a map of the flow, constructed by
observing it for specific times at which events occur in the
subsystems $\mathcal{X}_j$. An event is considered to happen when the
trajectory of one subsystem crosses a Poincar\'e section. So, given
two oscillators $S_1$ and $S_2$ (at least one being chaotic), with
trajectories in the subspaces where the phase is defined, the
conditional Poincar\'e map of $S_1$ is the trajectory position at the
moment that a series of equal events happens, in $S_2$.  Analogously,
the conditional Poincar\'e map of $S_2$ is the trajectory position at
the moment that a series of equal events happens, in $S_1$.  In the
case of periodically forced chaotic systems, an event may be defined
to happen whenever the forcing reaches a specific value, and the
conditional Poincar\'e map is the usual stroboscopic map, because the
time interval between two events is constant.  In coupled chaotic
systems, the time interval between two successive events is no longer
constant.

We define a time series of events $\tau^i_j$, by the following rule:
\begin{itemize}
\item {$\tau^i_1$} represents the time at which the $i$-th crossing of
  the trajectory of $S_2$ occurs in a Poincar\'e plane.
 
\item {$\tau^i_2$} represents the time at which the $i$-th crossing of
  the trajectory of $S_1$ occurs in a Poincar\'e plane.
\end{itemize}
\noindent
The discrete set of points observed at the times $\tau^i_j$ is called
set $\mathcal{D}$. This set, projected at the subspaces
$\mathcal{P}_j$ (where the phase is calculated), 
is named $\mathcal{D}_j$. The conditional Poincar\'e
map is represented by $T^{\tau_j^i}$. So, we say that $\mathcal{D}_j$
is the set of points generated by $T^{\tau_j^i}$.

The next step is to define when $\mathcal{D}_j$ can be regarded as
either a P-set or a PS-set, this last set implying phase
synchronization.

\section{Sets generated by the Conditional Poincar\'e Map}\label{P-sets}

The $\mathcal{D}_j$ set is a P-set, if it does not completely fulfill the
projection $\mathcal{X}_j$ of the attractor. In other words, a discrete
set $\mathcal{D}_j$ is considered to be a P-set, if for balls of
radius $\delta$, centered in all points of the attractor projection
$\mathcal{X}_j$, one does not find points of $\mathcal{D}_j$ inside of
all these balls. If $\mathcal{D}_j$ completely fulfill $\mathcal{X}_j$, we say
that these two sets are equivalent (and we represent this by the
symbol $\equiv$).  More details, see Appendix A.
  
So, a P-set exists, if the conditional Poincar\'e map is not transitive
on $\mathcal{X}_j$ \cite{transitivity}.  That is, the flow, observed
by the times for which the conditional Poincar\'e map is defined, does
not visit arbitrary regions of $\mathcal{X}_j$. Note that however, the
attractor is chaotic, and therefore, the chaotic set is always
transitive through the flow.  So, given a set of
initial conditions, its evolution through the flow eventually reaches
arbitrary open subsets of the original chaotic attractor.

We can classify three relevant types
of sets generated by the conditional Poincar\'e map 

\begin{description}
\item[type-a] $\mathcal{D}_j$ is equivalent to $\mathcal{X}_j$ 
($\mathcal{D}_j \equiv \mathcal{X}_j$). 
The conditional
  Poincar\'e map $T^{\tau_j^i}$ is transitive in $\mathcal{X}_j$. 
\item[P-set] \ $\mathcal{D}_j$ is NOT equivalent to $\mathcal{X}_j$
($\mathcal{D}_j \not\equiv \mathcal{X}_j$).
  The conditional Poincar\'e map is NOT transitive in $\mathcal{X}_j$.
\item[PS-set] \ $\mathcal{D}_j$ is P-set, with the additional
  condition that it is localized in the vicinity of the
  Poincar\'e section chosen to define the events.
\end{description}
\noindent

In the following, we comment each case: 

\subsection{type-a sets}

{\it These sets appear whenever there is no PS}.  
If two non-identical coupled chaotic systems
(topologically similar) are not phase synchronized, the chaotic
trajectories do not make correlated events in both subspaces
$\mathcal{P}_1$ and $\mathcal{P}_2$. As a consequence while the
trajectory is positioned at the specified Poincar\'e plane, at the
subspace $\mathcal{P}_1$, the trajectory in the subspace
$\mathcal{P}_2$ is everywhere in this subspace, making the set
$\mathcal{D}_j$ to be equivalent to $\mathcal{X}_j$.

An interesting illustration is the case of two uncoupled equal chaotic
systems, but with different initial conditions. As we construct the
conditional Poincar\'e maps, they will be a type-a set, since also the
distance between the trajectories in the two oscillators are
sensitive to the initial conditions, and will diverge exponentially.

\subsection{P-sets}

{\it These sets constitutes a necessary, but not sufficient, 
condition to describe PS.}

In periodically forced chaotic systems, they might appear when there
is not PS. As we already mentioned, the points in these sets are not
localized in special spots of the attractor projection.  As a
consequence, the domain of the absolute difference between the time at
which the same number of events happen in both oscillators has a broad
character.

\subsection{PS-sets}

{\it PS-sets imply PS and vice-versa}. They exist, if and only if,
there is phase synchronization, as shown in Appendix B, and
illustrated with the examples in Sec. \ref{sec:chua}, and throughout
this work. Another important point is that the PS-set provides a
real-time detection that can be easily experimentally implemented
(Sec.  \ref{sec:chua}), and easily constructed from a data set.

PS-set implies PS because the difference between the time at which the
$N$-th event happens in both oscillators is small, which means that
the time difference $|\tau_1^N - \tau_2^N|$ is smaller than a finite
constant value.  As a consequence, the points in the conditional
Poincar\'e map of one oscillator are confined around the Poincar\'e
section chosen to define the events. Therefore, the detection of a
PS-set can be done by observing this characteristic of the conditional
Poincar\'e map.

For R\"ossler-like oscillators, in which the trajectory spirals around
an equilibrium point, the PS-set is confined within an angular
section.
 
\subsection{Length of a PS-set}

Having found the PS-set, we can study properties of these sets that
give us the level of organization and coherence of the oscillators

A PS-set, $\mathcal{D}_j$, is said to have length 1, if the set is
constructed by the time series of the same events. Defining the event
to be given by the crossing of the trajectory to a Poincar\'e plane,
the corresponding time series of events $M_j(1)$ is given by
$\tau_j^{i},\tau_j^{i+1},\tau_j^{i+2},\tau_j^{i+3}, \ldots$.  For this
PS-set, points in $\mathcal{D}_j$ are mapped in $\mathcal{D}_j$,
after one application of the conditional Poincar\'e map.

A PS-set is said to have length 2, if it is obtained by a time series
of two different events.  So, a PS-set can be constructed from more
complex series of events.  We construct a length-2 basic set using a
time series of events $M_j(2)$ given by
$\tau_j^{i},\frac{\tau_j^i+\tau_j^{i+1}}{2},\tau_j^{i+1},\frac{\tau_j^{i+1}+
  \tau_j^{i+2}}{2},\ldots$ \cite{D2}.  As an example, for the
perturbed Chua's circuit, the length-2 basic set is constructed by a
stroboscopic map that collects points every half period of the
forcing. A length-2 basic set is assumed to be composed by two other
subsets, named minimal sets, the subsets $\mathcal{D}_j^0$ and
$\mathcal{D}_j^1$.

They have the property that if a point $x_0$ is such that $x_0 \in
\mathcal{D}_j^0$, this point, iterated by the conditional Poincar\'e
map, goes to the minimal set $\mathcal{D}_j^1$, and if $x_0$ is such
that $x_0 \in \mathcal{D}_j^1$, this point, iterated by the
conditional Poincar\'e map, goes to the minimal set $\mathcal{D}_j^1$.
Thus, points in $\mathcal{D}_j^1$ are mapped to itself after 2
applications of the conditional Poincar\'e map.  The minimal set
$\mathcal{D}_j^1$ is said to be disjoint to the set $\mathcal{D}_j^2$,
if they do not intersect, i.e., $\mathcal{D}_j^1 \cap \mathcal{D}_j^2 =
\emptyset$.  For some systems that present a strong phase-coherent
state, as the ones here studied, in which the instantaneous trajectory
velocity does not differ too much from the average velocity on typical
orbits, it is possible to find length-2 basic set, with disjoint
minimal sets, when PS is present.

For a general case, we do not expect to find a length-2 PS-set with
disjoint minimal sets. As an example, one can think of a spiking-firing
oscillator, phase synchronized with a  periodic forcing. Due to the fact that the
spiking-firing dynamics has a fast and a slow mode, the conditional
Poincar\'e maps might overlap.

\subsection{Sets diagram}

Here, we explain through a diagram the possible emerging sets from the
conditional Poincar\'e map.

$$
\xymatrix{
 & \mathcal{X}\ar[rd]^{\mathcal{P}_j} \ar[ld]_{\mathcal{P}_j \circ T} &  \\
\mathcal{D}_j \ar[rd] & &\mathcal{X}_j \ar[ld]\\
 &  \mathcal{D}_j \stackrel{?}{\equiv} \mathcal{X}_j \ar[ld]_{\txt{Yes}}
\ar[rd]^{\txt{No}} & \\
type-a &  & P-set \ar[d] \\
&   PS \ar@{<=>}[r]  &  PS-set
}
$$

Starting from the chaotic attractor $\mathcal{X}$, the set $\mathcal{D}$ is
constructed from the conditional Poincar\'e map, represented by $T$, 
we project $\mathcal{D}$ and $\mathcal{X}$ into the subspace
$\mathcal{P}_j$, obtaining the sets $\mathcal{D}_j$ and $\mathcal{X}_j$,
respectively.

We classify the set $\mathcal{D}_j$ into type-a set or P-set, by
checking whether the conditional Poincar\'e map is transitive in
$\mathcal{X}_j$, i.e., by verifying the equivalence between the sets
$\mathcal{X}_j$ and $\mathcal{D}_j$. Then, if the P-set is localized
in the vicinity of the Poincar\'e section where the events occur,
the P-set is a PS-set, which means that PS is present.

\section{PS-sets in the Chua's circuit}\label{sec:bsc}

\begin{figure}[!h]
  \centerline{\hbox{\psfig{file=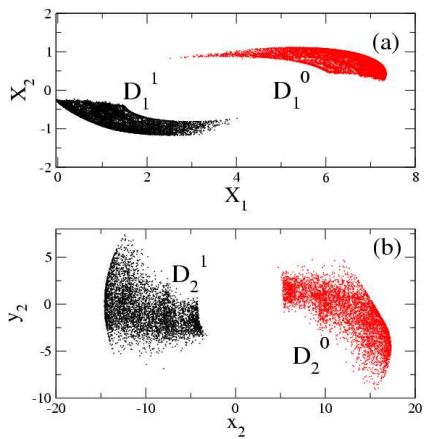,height=7cm}}}
  \caption{(a) Length-2 PS-set with disjoint minimal sets, in the
    forced Chua's circuit, for the parameters $f$ = 0.2801 and $V$ =
    0.001. The PS-set is constructed by the time series $M_1(2)$.
    (b) Length-2 PS-set with disjoint minimal sets, in the coupled
    R\"ossler oscillators, for the parameters $\epsilon$=0.01 and
    $\delta \alpha$ = 0.001. The PS-set is constructed by the time
    series $M_2(2)$.}
\label{PS_fig3}
\end{figure}

For applying our formalism to the periodically forced Chua's circuit,
the event times are $\tau_1^i = i\tau$, with $\tau$ representing the
forcing period.  The time series for the length-2 basic set is given
by ${1/2\tau,\tau,3/2\tau,2\tau,5/2\tau,3\tau,\ldots}$.  Everywhere
inside the $PS$ region, we find length-1 [as an example, see Fig.
\ref{PS_fig7}(a)] and length-2 [as an example, see Fig.
\ref{PS_fig3}(a)] {\bf PS}-sets, this latter with disjoint minimal
sets.  In Fig. \ref{PS_fig3}(a), the application of the conditional
Poincar\'e map in the minimal set $\mathcal{D}_1^0$ leads to the
minimal set $\mathcal{D}_1^1$. Both sets form a length-2 PS-set. Note
that both $\mathcal{D}_1^0$ and $\mathcal{D}_1^1$ can be regarded as a
length-1 PS-set.  To our numerical precision, we have checked that
there are not basic sets beneath the Synchronization region tip.
Which means that {\bf type-a} set is present for very small but finite
amplitude forcing. Outside the $PS$ region, there is a {\bf P-set},
i.e., a non-transitive conditional Poincar\'e map on the chaotic
attractor projection. So, $\mathcal{D}_j\not \equiv \mathcal{X}_j$. In
this case, there is not PS. More examples of length-Q PS-sets in the
phase synchronous forced Chua's circuit can be seen in \cite{firenze}.

\section{PS-sets in the coupled R\"ossler system} \label{sec:bsr}

We can use the formalism of the conditional Poincar\'e map to
study the appearance of PS in coupled chaotic systems, as the two
coupled R{\"{o}}ssler oscillators given by:

\begin{eqnarray}
\dot{x}_{1,2} = -\alpha_{1,2}y_{1}-z_{1}+\epsilon(x_{2,1}-x_{1,2})\\
\dot{y}_{1,2} =\alpha _{1,2}x_{1}+0.15y_{1}\\ 
\dot{ z}_{1,2} = 0.2+z_{1}(x_{1}-10)
\end{eqnarray}
\noindent with $\alpha _{1}=1$, and $\alpha _{2}=\alpha_{1}+\delta
\alpha $.  The index denotes systems 1 and 2. The subspaces
$\mathcal{P}_j$ are defined by $\mathcal{P}_j$=$(x_j,y_j)$.  In a
coupled chaotic system, $\tau_1^{i}$ ($\tau_2^{i}$) does not increase
uniformly, but it is given by the time the trajectory crosses the
Poincar\'e plane $y_{2}$=0 ($y_1$=0).  For these times, and using the
time series $M_j(2)$, we construct the minimal sets $\mathcal{D}_1^0$
and $\mathcal{D}_1^1$ ($\mathcal{D}_2^0$ and $\mathcal{D}_2^1$). The
parameters are $\epsilon$=0.01 and $\delta \alpha$ = 0.001.

In Fig.  \ref{PS_fig3}(b), we show a length-2 basic set with disjoint
minimal sets.  The application of the conditional Poincar\'e map in
the minimal set $\mathcal{D}_2^0$ leads to the minimal set
$\mathcal{D}_2^1$, and vice-versa. These two sets form together a
length-2 PS-set, but each one separated can be regarded as a length-1
PS-set. The characteristic of these PS-sets is that they appear as
localized structures around the Poincar\'e section chosen to define
the events. As one can see, the set $\mathcal{D}_2^0$ is localized in
the neighborhood of the line $y_2=0$, where the Poincar\'e section is
chosen.

\begin{figure}[!h]
\centerline{\hbox{\psfig{file=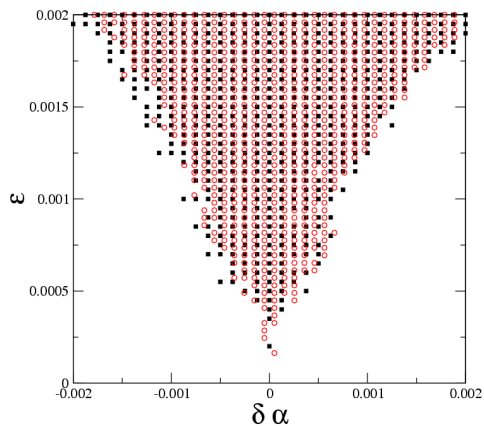,height=7cm}}} 
\caption{Empty  circles show parameters for which 
  PS exists, detected by Eq. (\ref{PS}), and filled squares represent
  parameters for which a length-1 PS-set appears simultaneously
  with a length-2 PS-set with disjoint minimal sets.  
Horizontal axis represents
  parameter mismatch and vertical axis, the coupling amplitude.}
\label{PS_fig4}
\end{figure}

In fact, this length-2 PS-set with disjoint minimal sets (as well as a
length-1 PS-set) is found everywhere in the PS region, as shown in
Fig. \ref{PS_fig4}. In it, filled squares represent parameters for
which these special PS-sets are found, and empty circles parameters
for which PS exists.

A PS-set of length-Q is detected using in Eq.
(\ref{practice_basic_sets}) $\mathcal{D}_2^{Q-1}$, from which we can
check whether $\mathcal{D}_2^{Q}$ occupies ({\bf type-a} discrete
set $\mathcal{D}$) the whole
space occupied by $\mathcal{X}_1$.  The set $B_{\ell}(x)$ in Eq.
(\ref{practice_basic_sets}) is constructed assuming squares of size
$\ell =1.5$, in points of the set $\mathcal{D}_2^{Q-1}$.  In Fig.
\ref{PS_fig4}, we show a case for $Q$=2.

\begin{figure}[!h]
  \centerline{\hbox{\psfig{file=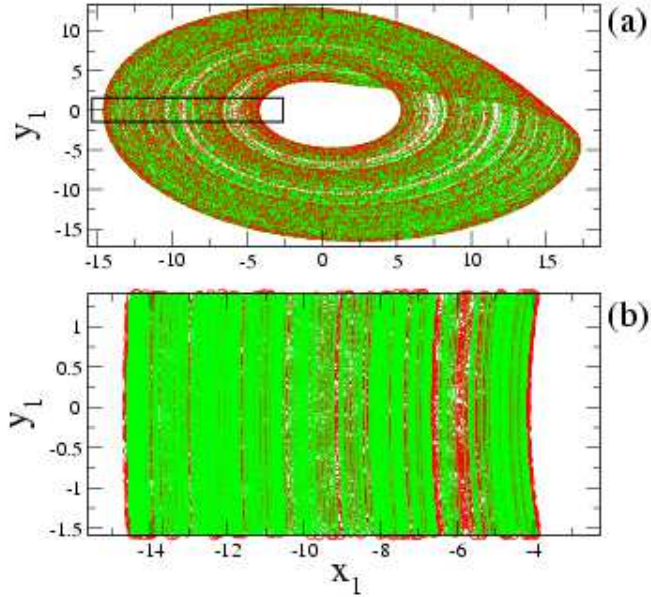,height=10cm}}}
  \caption{We show a situation where PS is not found for
    $\epsilon$=0.01 and $\delta \alpha$ = 0.00001. In (a), we show the
    attractor in the subspace $\mathcal{P}_1$, the subset
    $\mathcal{X}_1$ in gray, and the discrete set $\mathcal{D}_1$ in
    dark empty circles. In this figure, $\mathcal{D}_1$ is a type-a
    set, since $\mathcal{X}_1$ visits every neighborhood of points in
    $\mathcal{D}_1$ (the conditional Poincar\'e map is transitive in
    $\mathcal{X}_1$). In (b), we show a magnification of the box in
    (a).}
\label{PS_fig40}
\end{figure}

In Fig. \ref{PS_fig40}(a), we show the attractor in the subspace
$\mathcal{P}_1$, the subset $\mathcal{X}_1$ in gray, and the discrete
set $\mathcal{D}_1$ in dark empty circles . In (b), we show a
magnification of the box in (a).  Note that neighborhoods of arbitrary
points in the trajectory of $\mathcal{X}_1$ (gray) always contain a
point of the discrete set $\mathcal{D}_1$ (dark empty circles). So,
the set $\mathcal{X}_1$ is equivalent to the set $\mathcal{D}_1$ and,
therefore, $\mathcal{D}_1$ is not a PS-set, and therefore, it does not
exist PS.  Differently from the Chua's circuit, the R\"ossler coupled
system presents no P-sets for parameters outside of the PS region.

\section{Extension to other coupled chaotic systems}\label{extension}

The approach presented in this paper can be extended for non-coherent
attractors. As noticed in Ref. \cite{osipov}, attractors that present
non-coherent phase motion in the phase space may present a coherent
motion in the space of the velocities, i.e. $(\dot{x},\dot{y})$, which
is the case of the funnel attractor \cite{osipov}.  In this case, the
extension of our approach to the non-coherent phase motion is
straightforward. Instead of defining a conditional Poincar\'e map in
the phase space, we analyze the dynamics in the velocity space, in
which the phase is coherent, and therefore, all the ideas introduced
herein can be applied in this new space.

To some chaotic attractors, it might not be possible to define a
Poincar\'e section to construct the conditional Poincar\'e map.
However, one can still construct these conditional maps by defining
different types of events. For example, in coupled neurons, this event
can be chosen as being a beginning or the ending of the bursts, which
can be well defined by the crossing of the trajectory with some given
threshold.

\section{Conclusion and Remarks}\label{sec:cr}

A chaotic set is always transitive through the flow.
So, given a set of initial conditions, its evolution through the flow
eventually reaches arbitrary open subsets of the original chaotic
attractor.  However, a stroboscopic map of the flow, whose
generalization is here called conditional Poincar\'e map, might not
possess the transitive property. That is, given a set of initial
conditions, its evolution through the conditional Poincar\'e 
map {\bf \it might} not reach arbitrary open subsets of the chaotic attractor.

The introduction of the term ``conditional'' in the map nomenclature
comes from the unconventional and non rigidly way we adapt the
established definition of a stroboscopic Poincar\'e map. For coupled
chaotic oscillators, this conditional map is constructed based on
events, conveniently chosen at the same subspaces where the phase of a
chaotic system is defined. In addition, the application of this map
through the flow, which results in a discrete set, is inspected not in
the whole phase space, but also in the same subspaces where the phase
is defined.

If phase synchronization exists the conditional map generates
special discrete sets, named PS-sets.  The contrary is also true. i.e.,
a PS-set implies phase synchronization. This was illustrated in the
periodically forced Chua's circuit, in the coupled R\"ossler
oscillator, and in other more general (topologically equivalent)
coupled chaotic systems.

The ideas introduced here provide an efficient way of detecting phase
synchronization, without having actually to measure the phase of the
chaotic trajectory. Indeed, this detection can be done in experiments
in a real-time, as it was done here in the perturbed Chua's circuit.

It is worthy to say that the PS-sets are robust under small additive
noise that could corrupt the data in an experiment. This is so,
because a small additive noise does not interfere much with the time that
the trajectory crosses the Poincar\'e section, but just deviates the
timing of the crossing. Then, the PS-sets still remains. This
robustness against the noise is an important property in order to
apply these ideas in experiments, as done in this work.

We have also introduced the phase as a quantity that measures the
velocity of rotation of a projection of the tangent vector along the
trajectory.  This definition can be used to arbitrary flows,
independently whether or not they present a coherent or non-coherent
phase dynamics.

Finally, our formalism of the conditional Poincar\'e map can be used
in coupled maps (or perturbed) to detect synchronous behavior (not
{\bf phase} synchronization) between the systems (or between the
forcing), for the case where one does not find full synchronization
between the maps. One particular example where that happens is in the
periodically forced Logistic equation \cite{KLM} or for a system of
coupled Logistic maps \cite{shabunin}, where one can find a finite
number of synchronous chaotic subsets, the basic sets.

{\bf Acknowledgment} Research partially financed by FAPESP, Alexander
Von Humboldt Foundation, and CNPq.

\appendix\label{appA}

\section{The conditional Poincar\'e map and 
the $\mathcal{D}_j$ sets}

Next, we present some formalism using elements of Ergodic Theory
in order to introduce the conditional Poincar\'e map and the
$\mathcal{D}_j$ sets.

Given a flow $F_t$, we call $\mathcal{X}$ the chaotic attractor.  A
subset of the attractor is its projection on the subspace
$\mathcal{P}_j$, that we call $\mathcal{X}_j$. We assume $\mathcal{X}
\in R^d$.

The notion of stroboscopic map of a flow can be generalized to any
temporal translation on the trajectory. We define the temporal
translation to be a transformation represented by $T^{\tau^i_j}$,
called conditional Poincar\'e map. The initial condition $\vec{x}(t)$
is iterated under the temporal transformation $T^{\tau_j^i}$, to the
point $\vec{x}(t+\tau_j^i)$. Applying the transformation
$T^{\tau_j^i}$ in a typical trajectory, for the time sequence
$\tau_j^i$, $\tau_j^{i+1}$, $\tau_j^{i+2}$, give us the points
$\vec{x}(\tau_j^i)$, $\vec{x}(\tau_j^{i+1})$, and
$\vec{x}(\tau_j^{i+2})$.  So, $\vec{x}(\tau_j^{i+1})$ is the point
$\vec{x}(\tau_j^i)$ integrated by the flow for a time interval given
by $\tau_j^{i+1}-\tau_j^{i}$. Applying the transformation
$T^{\tau_j^i}$, for an infinite series of $\tau_j^i$, that is
$i={1,2,3,\ldots,\infty}$, gives us a discrete set that we call
$\mathcal{D}$. The projection of $\mathcal{D}$ on the subspace
$\mathcal{P}_j$ is named $\mathcal{D}_j$.

Now, we introduce the notion of transitivity.  Let us assume we have a
chaotic set $\mathcal{A}$. Let us choose two disjoint subsets
$\mathcal{B}$ and $\mathcal{C}$ in $\mathcal{A}$. So, $\mathcal{B}
\cap \mathcal{C} = \emptyset$.  There is a transformation $F$ that
generates the set $\mathcal{A}$, with the property that
$T(\mathcal{B})=\mathcal{C}$ and $F(\mathcal{C})=\mathcal{B}$. So,
clearly the transformation $F$ and its $n$-fold application ($F^n$)
can always place an arbitrary initial condition, belonging either in
$\mathcal{B}$ or $\mathcal{C}$, anywhere in the set $\mathcal{A}$.
This property makes the transformation $F$ to be transitive in the set
$\mathcal{A}$. However, the conditional Poincar\'e map might no longer
be transitive.  

For example, in the case of the $2n$-fold of the
transformation $T$, i.e., $T^{2n}$. Given an initial condition in the
set $\mathcal{B}$, its iteration by $T^{2n}$ will never reach the set
$\mathcal{C}$.  Therefore, the conditional Poincar\'e map $T^{2n}$ is not a
transitive transformation in the set $\mathcal{A}$. In the case of
flows, we have seen that the temporal transformation represented by
$T^{\tau_j^i}$, whenever PS is present, does not place arbitrary
points of the attractor everywhere in the subsets $\mathcal{X}_j$ of
the chaotic attractor. Therefore, if PS is present the transformation
$T^{\tau_j^i}$ is not transitive to this subset of the attractor.

The conditional Poincar\'e map $T^{\tau_j^i}$ is topologically
transitive in a set $\mathcal{A}$ \cite{wiggins} if for any two open
sets $\mathcal{B},\mathcal{C}\subset \mathcal{A}$, {\small
\begin{equation}
\exists \tau _j^{i}\text{ }/\text{ }T^{\tau_j{^i}}(B)\cap C\neq \emptyset. 
\label{transitive}
\end{equation}
\noindent} 
\noindent

To detect whether the conditional Poincar\'e map is transitive in some
subset, we introduce the notion of equivalence.  Two sets
$\mathcal{A}$ and $\mathcal{B}$ are (not) equivalent [equivalence is
represented by the symbol {\bf $\equiv$}, and non equivalence by {\bf
  $\not \equiv$}] if they (do not) occupy the same neighboring space.
In a more general way:

\begin{defi}
Two sets $\mathcal{A}$ and $\mathcal{B}$ are
equivalent, $\mathcal{A}\equiv \mathcal{B}$, if $\forall x \in
\mathcal{A}$, a set $\mathcal{C}$ can be constructed by the union of
open sets $B_{\ell }(x)$, open $R^{d}$ volumes centered at $x$ with
length $\ell $, such that $\forall y \in \mathcal{B}$ $\Longrightarrow
$ $y \in \mathcal{C}$, and $\mathcal{A}\not\equiv \mathcal{B}$ if $y
\notin \mathcal{C}$.
\end{defi}

\begin{defi}
 The set $\mathcal{D}$ is a P-set 
if $\mathcal{X}_j \not \equiv \mathcal{D}_j$. 
\end{defi}

\begin{defi}
If $\mathcal{D}_j$ can be decomposed into a
collection $\mathcal{D}_j={\mathcal{D}_j^{0},\mathcal{D}_j^{1},\ldots
  \mathcal{D}_j^{Q-1}}$ of subsets of $ \mathcal{X}_j$, with $Q\geq
1$, such that a point in $\mathcal{D}_j^{i}$ iterated by the 
conditional Poincar\'e map goes to $\mathcal{D}_j^{i+1(mod Q)}$, we refer to each
minimal set $\mathcal{D}_j^i$ as a \textit{recurrent decomposition}.
In the particular case we have $\tau_j^{i+1}-\tau_j^{i}=\tau$
(constant) this minimal set is a \textit{periodic decomposition}.  The
number of sets $Q$ is called \textit{length} of the decomposition
\cite{banks}.
\end{defi}

\begin{propo}
If $T^{\tau_j^{i}}$ is transitive on $\mathcal{X}_j$,
then $\mathcal{D}_j$ $\equiv \mathcal{X}_j $.
\end{propo}

\begin{propo}
If $T^{\tau_j^{i}}$ is non transitive in
$\mathcal{X}_j$, then a subset $\mathcal{A}$ can be constructed, such
that $\mathcal{A} \subset \mathcal{X}_j$ and $\mathcal{A}\not\equiv
\mathcal{X}_j$.
\end{propo}
Proofs of Propositions 1 and 2 can be done by using Eq.
(\ref{transitive}) and the definitions.

Definition 3 can be understood by the PS-set 
of length-1 and length-2 considered in this work.  If the
length-1 PS-set is constructed by the time series
$\tau_j^{i},\tau_j^{i+1},\tau_j^{i+2},\tau_j^{i+3}, \ldots$, the
length-2 is obtained by the time series given by
$\tau_j^{i},\frac{\tau_j^i+\tau_j^{i+1}}{2},\tau_j^{i+1},\frac{\tau_j^{i+1}+
  \tau_j^{i+2}}{2},\ldots$ \cite{D2}.  For the particular case of a
periodically forced system, $\tau_1^{i+1}-\tau_1^{i}$=$1/f$, and so,
the length-1 PS-set is the standard stroboscopic map, that collects
point every period of the forcing. The length-2 PS-set is
constructed by a stroboscopic map that collects points every half
period of the forcing. For the length-2 PS-set, we have two minimal
sets, the sets $\mathcal{D}_j^0$ and $\mathcal{D}_j^1$, with the
property that if $x_0 \in \mathcal{D}_j^0$, this point iterated by the
conditional Poincar\'e map goes to $\mathcal{D}_j^1$ and $\mathcal{D}_j^1$
goes to $\mathcal{D}_j^0$, under the 
conditional Poincar\'e map. 

To check whether $\mathcal{D}_j \not\equiv \mathcal{X}_j$, 
we do the following, for $x \in \mathcal{X}_j$ it exists $y \in
\mathcal{D}_j$ such that 
\begin{equation}
y \cap B_{\ell}(x) = \emptyset,  
\label{practice_basic_sets}
\end{equation}
\noindent
where $B_{\ell}(x)$ is a open ball of radius $\delta$ centered at the
point $x$. $\delta$ is a small positive value.

\section{The PS-sets}\label{appB}

In this appendices, we show that a PS-set exists if, and only if, phase
synchronization exists, which in other words PS-set implies PS and PS
implies PS-sets.

Given a dynamical system ${\bf Y}^{\prime} = {G}({\bf Y})$, let
$F^t$ be the flow and $\mathcal{X}$ the attractor generated by it, we
suppose that we have a chaotic dynamics now on.  Let $\Sigma_j$ be the
Poincar\'e section in the subspace $\mathcal{P}_j$, and let $\Pi_j$ be
the Poincar\'e map associated to the section $\Sigma_j$, such that
given a point $x_j^{i} \in \Sigma_j$, thus $x_j^{i+1} =
\Pi_j(x_j^{i})$ = $F^{\Delta \tau _j ^{i+1}}(x_j^i)$.  From now on we use a rescaled time
$t^{\prime} = t/ \langle T_1 \rangle $.  For a slight abuse of
notation we omit the symbol $\prime$.

\begin{propo}
  Given two interacting oscillators. Then PS-sets can be constructed
  if, and only if, phase synchronization is present.
\end{propo}

To show the {\it if} in the preposition let us start by considering
the time interval associated to the return of the point $x_j^{i}$ to
the point $x_j^{i+1}$ is $\Delta \tau_j^{i+1}$ = $\tau_j^{i+1} -
\tau_j^{i}$, with $\tau_j^{i+1}$ being the times at which the
subsystem $\mathcal{X}_j$ crosses the Poincar\'e section $\Sigma_j$,
in the subspace $\mathcal{P}_j$.

As already introduced, the average return time is given by $\langle
T_j \rangle = \frac{\sum_{i=0}^{N} \Delta \tau^j_1 }{N} =
\frac{\tau_j^N}{N}$, and the time is rescaled, such that $\langle T_1
\rangle =1$. Our hypothesis is that the subsystem $\mathcal{X}_j$ has
a phase-coherent oscillation, so there is a number $\delta_j$ for
which holds \cite{kreso}:
\begin{equation}
|  \tau _j ^{N} - N | \leq \delta_j .
\label{eqB1}
\end{equation}
\noindent
The number $\delta_j < 1$ measures the coherence in
the phase oscillation, and is linked to the phase diffusion
\cite{reviews,kreso}. This equation holds for all $N$, so it implies
that for a single oscillation is also true that $| \Delta \tau_j ^{i} -
\langle T_j \rangle | \leq \delta_j$. In the case of two systems that
present PS, it holds \cite{murilo_potsdam}:

\begin{equation}
\Big|  \tau _1 ^{N} -  \tau _2 ^{N} \Big| \leq \delta_3,
\label{diferenca_temporal}
\end{equation}
\noindent
with $\delta_3 < 1$. This equation implies $| \Delta \tau_1 ^{i} -
\Delta \tau_2 ^{i}| \leq \delta_3$, which states that the time
intervals in a single oscillation are strongly related in phase
synchronization.

Let us introduce a new variable that measures the difference between
the time interval of two events in $\mathcal{X}_1$ and
$\mathcal{X}_2$. This new variable is $\Delta \tau _{2,1}^i = \Delta
\tau_2^i - \Delta \tau_1^i$, from Eq. 
(\ref{diferenca_temporal}), it is true
that $| \Delta \tau _{2,1}^i | \leq \delta_3$.

Now, we analyze one typical oscillation. Given the following initial
conditions $x_1^0 \in \Sigma_1$ and $x_2^0 \in \Sigma_2$, we evolve
both until $x_1^0$ returns to $\Sigma_1$. In other words, we evolve
both initial conditions for a time $\Delta \tau_2 ^ 1 $. So,
$F^{\Delta \tau _2 ^{1}}(x_1^0) = \Pi_2(x_1^0)= x_1^1 \in \Sigma_1 $.
Analogously, $F^{\Delta \tau _2 ^{1}}(x_2^0) = F^{ \Delta \tau _1 ^{1}
  + \Delta \tau _{2,1}^1}(x_2^0) = F^{ \Delta \tau _{2,1}^1} \circ
F^{ \Delta \tau _1 ^{1}}(x_2^0) = F^{ \Delta \tau _{2,1}^1}(x_2^1)$.

Now, we use the fact that $| \Delta \tau _{2,1}^i | < \delta_3 $, and
write that

\begin{equation}
 F^{ \Delta \tau _{2,1}^1}(x_2^1) \approx x_2^1 + {\bf G}(x_2^1) \delta_3.
 \label{tiagoIII}
\end{equation}
\noindent
So, given an initial condition in $\Sigma_1$ (subsystem
$\mathcal{X}_1$) evaluated by the time initial conditions return in
the section $\Sigma_2$ (of subsystem $\mathcal{X}_2$), it returns near
the section $\Sigma_1$, and vice-versa.

For a general case, we have to show that an initial condition, on the
section $\Sigma_1$, evolved by the flow for the time $\sum_{i=0}^N
\Delta \tau_{2,1}^i$ still remains close to this section. In other
words, we have to show that the approximation in Eq. (\ref{tiagoIII})
is valid for an arbitrary number of events $N$ in the subspace
$\mathcal{X}_2$.  Now, noting that $\sum_{i=0}^N \Delta \tau_{2,1}^i =
\tau_{2}^N - \tau_1^N$, from our hypotheses of phase coherent dynamics
$|\sum_{i=0}^N \Delta \tau_{2,1}^N| = |\tau2^N - \tau_1^N | <
\delta_3$. The same arguments used to derive Eq. (\ref{tiagoIII}) for
one oscillation, can be extended to an arbitrary number $N$, so we proof
the {\it if} of the proposition (PS implies the existence of PS-sets).

Now, to show the {\it only if} (PS-sets imply PS), let us say that 
there is a PS-set.  As a consequence, Eq. (\ref{diferenca_temporal})
is valid.  Then using the definition of phase coherence \cite{kreso},
and noting that in PS the average return times in a given Poincar\'e
section are the same, we see that Eq. (\ref{diferenca_temporal}) comes
from the boundness of the phase, concluding the proposition.

Furthermore, let us suppose that the trajectory of the oscillators are
perturbed by a small perturbation, which does not destroy the phase
synchronous dynamics. The effect of the small perturbation in the time
return of the trajectory to a Poincar\'e section is to deviate this
time according $\bar T_j^i = T_j^i + \xi_j^i$, where $\xi_j^i$
represents the perturbation in system $S_j$ at the moment of the i-th
event, with $max_i|\xi^i_j|< \kappa$. Under these hypotheses about the
perturbation, we conclude the following result.

\begin{propo}
The PS-set is robust under perturbations
\end{propo}

This result shows that a PS-set can be constructed in PS states.
Moreover, for the coupled R\"ossler-like systems, this result states
that this set is confined in an angular region, which is a consequence
of Eq. (\ref{diferenca_temporal}).

To see the relation between the constant $\delta_3$ and the size of
the PS-set, we do the following. By the time
normalization, average time interval between points in phase space are
proportional to their distances. So, from Eq. (\ref{tiagoIII}) we write
$\delta_3 = |\mathcal{H}| / {\bf G}(x_2^0) $, with $\mathcal{H}$ being
the average half length of the PS-set.  A rough calculation shows
that in our experiment with the Chua's circuit, we have that $\delta_3
\approx 1 / 2.5$, and for the coupled R\"ossler oscillators, we have
that $\delta_3 \approx 1 / 2.1$. This results completely agree with
the theoretical approach done in \cite{murilo_potsdam}.


\begin{thebibliography}{99}
\bibitem{phase_definitions} phase can be defined as a Hilbert
  transformation of a trajectory component, as the angle of a
  trajectory point into a special projection of the attractor, as a
  function that grows $2\pi$ every time the chaotic trajectory crosses
  some specific surface, or as the angle of the projected vector field
  into some subspace with respect to some rotation point.
\bibitem{rossler} The results presented here for the sinusoidally 
forced Chua's circuit were also verified in the sinusoidally 
forced R\"ossler oscillator.
\bibitem{cita_r} $r$ is considered to be a rational.
  However, as shown in Ref. \cite{murilo}, PS, as defined by the
  boundness of the phase difference was found in two chaotic systems
  for a finite but very large time interval, as $r$ approaches an
  irrational.  Therefore, although in this work we consider $r$ to be
  rational, we should make the remark that for the especial situation
  as the one presented in Ref.  \cite{murilo}, Eq. (\ref{PS}) can only
  be satisfied for a finite but large time if $r$ is considered to be
  irrational.
\bibitem{intermittency} The intermittency observed in the phase
  difference is characterized as an usual intermittency, by the
  alternance between a laminar regime and a burst regime. If the phase
  difference remains bounded in the interval $[0,\langle \Delta \phi_1
  \rangle]$, we say that we have a laminar regime. If the phase
  difference leaves this interval, we have a burst, also known as
  phase slip. As one approaches the border between the PS and the non
  PS region, the laminar regime in the phase difference becomes
  longer.
\bibitem{D2} The choice of these time intervals is not unique. It
  depends on what type of event one wants to identify in the system. A
  length-1 basic set is constructed by measuring time intervals
  between the occurrence of two events of the same type. $A$
  length-2 basic set is constructed by measuring the time interval
  between the occurrence of an event $A$ and an event $B$, and then,
  between the event $B$ and finally the event $A$, and so, on.
\end{thebibliography}
\end{document}